\documentclass[aps,twocolumn,epsfig,graphics,floatfix, mathbbm,pra]{revtex4}

\usepackage{amsmath,amsfonts,amssymb,graphics,graphicx,epsfig,color,times,bbm}

\def\r{\rangle}
\newcommand{\id}{{\mathbbm{1}}}

\begin{document}
\bibliographystyle{apsrev}

\title{Resilience of multi-photon entanglement under losses}

\author{G.A.\ Durkin$^{1,2}$}
\email{gabriel.durkin@qubit.org}

\author{C.\  Simon$^{1,2,3}$}

\author{J.\ Eisert$^{4}$}

\author{D.\ Bouwmeester$^2$}

\affiliation{
$^1$ Centre for Quantum Computation, University of Oxford, OX1 3PU, UK \\
$^2$  Department of Physics, University of California, Santa
Barbara, CA 93106, USA \\
$^3$ Laboratoire de Spectrom\'{e}trie
Physique, CNRS et Universit\'{e} J.\ Fourier - Grenoble,
BP 87, 38402 St.\ Martin d'H\`{e}res, France\\
$^4$ Institut f\"{u}r Physik, University of
Potsdam,
D-14469 Potsdam, Germany
}

\date{\today}

\begin{abstract}
We analyze the resilience under photon loss of the bi-partite 
entanglement present in multi-photon states produced by parametric
down-conversion. The quantification of the entanglement is made
possible by a symmetry of the states that persists even under
polarization-independent losses. We examine the approach of the
states to the set of  states with a positive partial transpose
as losses increase, and calculate
the relative entropy of entanglement. We find that some
bi-partite distillable entanglement persists for arbitrarily high losses.
\end{abstract}


\maketitle

\section{Introduction}

Parametric down-conversion has been used in many experiments
\cite{pdcexp} to create polarization entangled photon pairs
\cite{kwiat}. Recent experimental \cite{lamasdemart,eisenberg} and
theoretical \cite{durkin,simon,reid} work has studied the creation
of strong entanglement of large numbers of photons. The states
under consideration are entangled pairs of light pulses such that
the polarization of each pulse is completely undetermined, but the
polarizations of the two pulses are always anti-correlated. Such
states are the polarization equivalent of approximate singlet
states of two potentially very large spins \cite{howell}. An
application of the states for quantum key distribution has been
suggested \cite{durkin}.

In any realistic experiment photons will be lost
during 
propagation.
It is therefore of great practical interest to analyze
the resilience of the multi-photon entanglement under loss. A
priori this seems like a very difficult task, because it requires
the quantification of the entanglement present in mixed quantum
states of high or actually even infinite dimensionality. However, the
multi-photon states introduced in the above work exhibit very high
symmetry - in the absence of losses they are spin singlets. The
related symmetry under joint polarization transformations on both
pulses is preserved even in the presence of polarization-independent
losses. This makes it possible to apply the concepts of
`entanglement under symmetry' developed in Refs.\
\cite{Werner,Rains,Sym,Regul,KGH} to the quantification of the
multi-photon entanglement in the presence of losses. 
We calculate the degree of entanglement for the 
resulting states of high symmetry, as quantified in terms of the 
relative entropy of entanglement. We  show that
some (distillable) entanglement remains for arbitrarily high losses.

\section{Symmetry of the states in the presence of losses}

In the above-mentioned experiments and proposals a non-linear
crystal is pumped with a strong laser pulse, and a three-wave
mixing effect leads to the creation of photons
along two directions $a$ and $b$.   To a good
approximation the Hamiltonian 
in the interaction picture in a four-mode description
is given by
\begin{equation}\label{Hamiltonian}
    H=e^{i\phi}
    \kappa({a}_{h}^{\dagger }
    {b}_{v}^{\dagger}-{a}_{v}^{\dagger }{b}_{h}^{\dagger}) +
    e^{- i\phi}
    \kappa({a}_{h}{b}_{v}-{a}_{v}{b}_{h}).
\end{equation}
The real coupling constant $\kappa$ is proportional to the
amplitude of the pump field and to the relevant non-linear optical
coefficient of the crystal, and $\phi$ denotes the phase of the
pump field. Photons are created into the four modes with annihilation
operators $a_{h}$,
$a_{v}$, $b_{h}$, $b_{v}$, where $h$ and $v$ denote horizontal and
vertical polarization. Note that both the modes and the associated
annihilation operators will be denoted with the same symbol. In
the absence of losses, this Hamiltonian leads to a state vector of the
form \cite{durkin,simon}
\begin{eqnarray}
|\psi\rangle = e^{-iHt}|0\rangle= \frac{1}{\cosh^{2}\tau}\sum
_{n=0}^{\infty}e^{i n \phi} \sqrt{n+1} \;\tanh^{n}\tau\;|\psi^{n}_{-}\rangle,
\label{pdcstate}
\end{eqnarray}
where $\tau=\kappa t$ is the effective interaction time and
\begin{eqnarray}
&&|\psi^{n}_{-}\rangle = \frac{1}{\sqrt{n+1}}
\frac{1}{n!}(a^{\dagger }_{h}b^{\dagger
}_{v}-a^{\dagger }_{v}b^{\dagger }_{h})^n |0\rangle\label{psin-}\\
&&= \frac{1}{\sqrt{n+1}}\sum_{m=0}^{n}(-1)^{m}|n\!\!-\!\!m\r_{a_h}
|\,m\r_{a_v}|\,m\r_{b_h}|n\!\!-\!\!m\rangle_{b_v} \,. \nonumber
\end{eqnarray}
In experiments the pump phase is typically unknown, and data is
collected over time intervals much longer than the pump field
coherence time. We will therefore consider the state $\rho$
obtained from the state vector Eq. (\ref{pdcstate}) by uniformly
averaging over the pump phase $\phi\in[0,2\pi)$:
\begin{equation}
\rho = \frac{1}{\cosh^{4}\tau}\sum _{n=0}^{\infty}(n+1)
\;\tanh^{2n}\tau\;|\psi^{n}_{-}\rangle \langle \psi^{n}_{-}|
\label{pdcrho}.
\end{equation}
The Hamiltonian $H$ is invariant under any joint polarization
transformation in the spatial modes $a$ and $b$. That is, if one
defines ${\bf a}=(a_h,a_v)$ and ${\bf b}=(b_h,b_v)$, then $H$ is
invariant under the joint application of the same unitary $U$ from
$SU(2)$ to both vectors, ${\bf a} \mapsto {U}{\bf a}$ and ${\bf b}
\mapsto { U} {\bf b}$. This invariance of $H$ is inherited by the
multi-photon states created through the action of $H$ on the
vacuum. This symmetry can be expressed as
\begin{equation}
V(U)\rho V(U)^{\dagger}=\rho \label{sym}
\end{equation}
for all $U\in SU(2)$, where $V(U)= e^{i {\bf n} {\bf J}}$, and the
real vector ${\bf n}$ is specified by $U=e^{i{\bf n}\sigma/2}$,
$\sigma$ denoting the vector of Pauli matrices. Here  the angular
momentum operator  ${\bf J}$ can be written as ${\bf J}={\bf
J}_{a}+ {\bf J}_{b}$. The components of ${\bf J}_{a}$ associated
with spatial mode $a$ are given by the familiar quantum Stokes
parameters ${J}_{a,x}=(a_{+}^{\dagger} a_{+} - a_{-}^{\dagger}
a_{-})/2$, ${J}_{a,y}=(a_{l}^{\dagger} a_{l} - a_{r}^{\dagger}
a_{r})/2$, and
${J}_{a,z}=(a_{h}^{\dagger} a_{h} - a_{v}^{\dagger}
a_{v})/2$, with $a_{\pm}=(a_{h}\pm a_{v})/\sqrt{2}$ corresponding
to light that is linearly polarized at $\pm 45^o$, and
$a_{l,r}=(a_{h}\pm i a_{v})/\sqrt{2}$ to left and right-hand
circularly polarized light.  Analogous relations hold for spatial
mode $b$.

In the present work we are interested in the states created by $H$
in the presence of losses. These losses will be modeled by four
beam splitters of transmittivity $\eta\in [0,1]$, one for each of
the modes $a_h, a_v, b_h, b_v$, where the modes are mixed with
vacuum modes. Explicitly, the operation ${\cal L}^a_{\eta}$
corresponding to losses characterized by $\eta$ acting on a single
mode $a$ is given by
\begin{equation} {\cal L}^a_{\eta}(\rho)=\sum
\limits_{n=0}^{\infty} L_n^a \rho (L_n^a)^{\dagger},
\end{equation}
with $L_n^a$ being given by
\begin{equation}
L_n^a=\frac{1}{\sqrt{n!}}(1-\eta)^{\frac{n}{2}} a^n
\eta^{\frac{1}{2}a^{\dagger}a}.
\end{equation}
One can easily verify that these operators satisfy
\begin{equation}
\sum \limits_{n=0}^{\infty} (L_n^a)^{\dagger} L_n^a = \openone,
\end{equation}
required for trace preservation. In this paper we are interested
in the situation where an equal amount of loss occurs in all four
modes. We will denote the corresponding quantum operation by
\begin{equation}
{\cal L}_{\eta}={\cal L}^{a_h}_{\eta}\otimes {\cal
L}^{a_v}_{\eta}\otimes{\cal L}^{b_h}_{\eta}\otimes{\cal
L}^{b_v}_{\eta}. \label{loss}
\end{equation}
It is not difficult to apply this loss channel to the state $\rho$
of Eq. (\ref{pdcrho}). However, the resulting expression is quite
unwieldy, and quantifying the entanglement present in the state
seems like a hopeless task at first sight. We will now discuss
general properties of the resulting state that allow a simple
parametrization and as a consequence the determination of its
entanglement.

In the absence of losses, all components of the state created by
the action of $H$  have an equal number of photons in the $a$
modes and in the $b$ modes, since photons are created in pairs.
The state vector $|\psi\rangle$ of Eq. (\ref{pdcstate}) is a
superposition of terms corresponding to different total photon
numbers. For any given term we will denote the number of photons
in the $a$ modes by $\alpha=\alpha_h+\alpha_v$, where $\alpha_h$
is the number of photons in mode $a_h$ etc. Analogously, the
number of photons in the $b$ modes is denoted by
$\beta=\beta_h+\beta_v$. The relative phase between terms with
different values of $\alpha$ or $\beta$ depends on the pump phase
$\phi$.  The corresponding coherences in the density matrix are
removed when averaging over the pump phase.

Losses  lead to the appearance of terms with $\alpha \neq \beta$.
The state $\rho'= {\cal L}_{\eta}(\rho)$ after losses now has the
form
\begin{equation}
    \rho'=\sum_{\alpha,\beta=0}^{\infty} P{(\alpha,\beta)}
    \rho^{(\alpha,\beta)}, \label{rho} \end{equation} where
$P{(\alpha,\beta)}$ is the probability to have photon numbers
$\alpha$ and $\beta$ in the $a$ and $b$ modes respectively, and
$\rho^{(\alpha,\beta)}$ is the corresponding state. In the state
before losses, the terms $\rho^{(\alpha,\alpha)}$ are maximally
entangled states (for $\alpha \neq 0$), denoted by
$|\psi_-^{\alpha}\rangle \langle \psi_-^{\alpha}|$ in the notation
of Eq. (\ref{psin-}). Losses reduce this entanglement, but do not
make the state become separable, as will be seen below.

The state vector
$|\alpha_h,\alpha_v\rangle|\beta_h,\beta_v\rangle$ corresponds to
a spin state vector $|j_a,m_a\rangle|j_b,m_b\rangle$ with
$j_a=(\alpha_h+\alpha_v)/2,m_a=(\alpha_h-\alpha_v)/2,j_b=(\beta_h+\beta_v)/2,m_b=(\beta_h-\beta_v)/2$.
Note that in this representation a single photon corresponds to a
spin-1/2 system. A state with fixed photon numbers $\alpha$ and
$\beta$ thus corresponds to a state of two fixed general spins
$j_a=\alpha/2$ and $j_b=\beta/2$.

The key feature of the lossy channel ${\cal L}_{\eta}$ of Eq.
(\ref{loss}) is that it does not destroy the symmetry described by
Eq. (\ref{sym}). We have that
\begin{equation}
V(U) {\cal L}_{\eta}(\rho)V(U)^{\dagger}={\cal L}_{\eta}(\rho)
\end{equation}
for all losses $\eta$ and all $U\in SU(2)$. To sketch the argument
why this symmetry is retained we will resort to the Heisenberg
picture. Polarisation-independent loss in the $a$ modes can  be
described by the map
\begin{equation} {\bf a}\mapsto {\bf
a'}=\sqrt{\eta}{\bf a}+\sqrt{1-\eta} {\bf c},
\end{equation}
where ${\bf c}=(c_h,c_v)$ is a vector of unpopulated modes that are coupled
into the system due to the loss. Applying ${ U}\in SU(2)$ to ${\bf
a'}$ gives
\begin{equation}
{\bf a''}={ U}{\bf a'}=\sqrt{\eta}{ U}{\bf a}+\sqrt{1-\eta}{
U}{\bf c}.
\end{equation}
On the other hand, applying first ${ U}$ and then the loss
operation gives
\begin{equation}
{\bf a''}=\sqrt{\eta}{ U}{\bf a}+\sqrt{1-\eta}{\bf c},
\end{equation}
in which the last term is different. 
However, this term just
corresponds to a coupling in of unpopulated modes with a
coefficient $\sqrt{1-\eta}$. The resulting lossy channel is
invariant under the map ${\bf c}\mapsto U{\bf c}$, since these modes
are unpopulated. This implies
that the state after application of the loss operation ${\cal
L}_{\eta}$ has the same symmetry as before. 
Note that for this
argument to hold, the amount of loss in the $a$ and $b$ modes does
not have to be the same, since
the transformations are applied independently to each of 
$\mathbf{a}$ and $\mathbf{b}$. However, within each
spatial mode, losses must be polarisation insensitive.

The identification of the above symmetry dramatically
simplifies the description of the resulting states.
The most general state $\rho^{(\alpha,\beta)}$ with
fixed value of $\alpha$ and $\beta$ for which
$V(U)\rho^{(\alpha,\beta)}V(U)^{\dagger}= \rho^{(\alpha,\beta)}$
for all $U\in SU(2)$ is of the form
\begin{equation} \label{rhoalphabeta}
\rho^{(\alpha,\beta)}=\sum_{j=|j_a-j_b|}^{j_a+j_b}
\mu^{(\alpha,\beta)}_j\Omega^{(\alpha,\beta)}_j,
\end{equation}
where  $j_a=\alpha/2,j_b=\beta/2$ \cite{KGH}, essentially
as a consequence of Schur's lemma \cite{grouptheory}.
Here, the
$\mu^{(\alpha,\beta)}_j$ form a probability distribution for all
$(\alpha,\beta)$ in the allowed values for $j$.
In turn, $\Omega^{(\alpha,\beta)}_j$ is up to normalization to unit trace
a projection onto the space of total spin $j$ (for fixed $j_a=\alpha/2,j_b=\beta/2$).
That is, $\Omega^{(\alpha,\beta)}_j=
\id^{(\alpha,\beta)}_j/(2j+1)$, where $\id^{(\alpha,\beta)}_j$ is
equal to the identity when acting on the space labeled
by $\alpha$, $\beta$, and $j$, and zero otherwise \cite{KGH,Schliemann}.

As an example, let us consider the case with exactly one photon in
each spatial mode, i.e., $\alpha=\beta=1$. Then there are just two terms
in the expansion of Eq.\ (\ref{rhoalphabeta}), proportional to
$\Omega^{(1,1)}_0$ and $\Omega^{(1,1)}_1$. The state
$\Omega^{(1,1)}_0$ is the projector onto the two-photon singlet
state with state vector
$(({a}_{h}^{\dagger }
{b}_{v}^{\dagger}-{a}_{v}^{\dagger}{b}_{h}^{\dagger})/\sqrt{2})|0\rangle$,
while $\Omega^{(1,1)}_1$ is the normalized projector onto the
spin-1 triplet. The trace condition
$\mu^{(1,1)}_{0}+\mu^{(1,1)}_{1}=1$ means that the set of all
invariant states $\rho^{(1,1)}$ is characterized by just one
parameter. Note that the most general state with exactly one
photon in each spatial mode would be characterized by $15$
parameters.

\section{Quantifying the entanglement}

In order to quantify the entanglement in a given physical
situation, one has to determine the coefficients
$P{(\alpha,\beta)}$ of Eq.\ (\ref{rho}) and
$\mu^{(\alpha,\beta)}_j$ of Eq.\ (\ref{rhoalphabeta}), which may
be calculated from the polarization dependent photon counting
probabilities $p(\alpha_h,\alpha_v,\beta_h,\beta_v)$. These in
turn can be determined by explicitly applying the loss channel
${\cal L}_{\eta}$ of Eq. (\ref{loss}) to the state $\rho$ of Eq.
(\ref{pdcrho}). One finds
\begin{eqnarray}
&&p(\alpha_h,\alpha_v,\beta_h,\beta_v)=\frac{
\eta^{\alpha+\beta}(1-\eta)^{\alpha+\beta}}{(\cosh (\kappa t))^4
\alpha_h! \alpha_v! \beta_h!
\beta_v!}\label{probs}
 \\
&&\times \sum \limits_{m= m_{0},n=n_{0}}^{\infty}
\frac{((1-\eta)\tanh (\kappa
t))^{2(m+n)}(m!)^2
(n!)^2}{(m-\alpha_h)!(m-\beta_v)!(n-\alpha_v)!(n-\beta_h)!},\nonumber
\end{eqnarray}
where $m_0=\max(\alpha_h,\beta_v)$ and
$n_{0}=\max(\alpha_v,\beta_h)$. The probabilities
$P(\alpha,\beta)$ are obtained by summing this expression over all
$\alpha_h,\alpha_v,\beta_h,\beta_v$
with
$\alpha_h+\alpha_v=\alpha$ and $\beta_h+\beta_v=\beta$.

The coefficients $\mu^{(\alpha,\beta)}_j$ may be written as linear combinations of
the $p(\alpha_h,\alpha_v,\beta_h,\beta_v)$ via the Clebsch-Gordan
coefficients \cite{grouptheory}  by means
of the standard procedure of `coupling spins'. Polarization-sensitive
photon counting in the spatial modes $a$
and $b$ corresponds to the basis spanned by the
$|j_a,m_a\rangle|j_b,m_b\rangle$, while the
$\mu^{(\alpha,\beta)}_j$ and $\Omega^{(\alpha,\beta)}_j$ are
defined in terms of the total spin, corresponding to the label $j$. 
Since the $\mu^{(\alpha,\beta)}_j$  characterize the normalized state
$\rho^{(\alpha,\beta)}$, they only depend on the relative
probabilities of the different values of
$\alpha_h,\alpha_v,\beta_h,\beta_v$ for given $\alpha$ and
$\beta$. Eq.\ (\ref{probs}) then implies that they depend on the
interaction time $t$ and the transmission $\eta$ only via the
combination $\xi=(1-\eta)\tanh(\kappa t)\in [0,1]$, which ranges
from zero for perfect transmission (or, less interestingly, zero
interaction time) to one in  a
limit of complete loss and
infinite interaction time.

For example, for $\alpha=\beta=1$, the single independent
parameter $\mu^{(1,1)}_0$ is given by
\begin{eqnarray}
\mu^{(1,1)}_{0}= 1 &-& \frac{3}{2}  \bigl(
 p(1,0,1,0)+p(0,1,0,1)\bigr)/P(1,1).\nonumber\\
 \end{eqnarray}
where as before $P(1,1)=p(1,0,1,0)+p(1,0,0,1)+p(0,1,1,0)+p(0,1,0,1)$.
This gives
\begin{equation}
    \mu^{(1,1)}_{0}=(1+ \xi^2/2)/(1+2\xi^2).\label{muzero}
\end{equation}

To quantify the entanglement present in the total state, one can
proceed by considering each $\rho^{(\alpha,\beta)}$ separately.
There is no unique measure of entanglement for mixed states.
Instead, there are several inequivalent ones, each of which is
associated with a different physical operational interpretation \cite{Int}.
The relative entropy of entanglement \cite{Relent}, which will be
employed in the present paper 
specifies to which extent a given state can be
operationally distinguished from the closest state that is
regarded as being disentangled. The relative entropy of
entanglement of a state $\rho$ is defined as
\begin{eqnarray}\label{relent}
E_{R}(\rho )= \inf_{\sigma \in \mathcal{D}}
  S(\rho||\sigma),
\end{eqnarray}
where $S(\rho||\sigma)=\text{tr} [\rho \log \rho - \rho \log
\sigma]$ denotes the quantum relative entropy of the state $\rho$
relative to the state $\sigma$.  Here ${\mathcal D}$ is taken
to be the set of states with positive partial transpose
\cite{peres} (PPT states). This set of states includes the set of
separable states, but in general also contains bound entangled
states \cite{boundent}. The relative entropy of entanglement
is an upper bound to the distillable entanglement \cite{Int}, 
providing a measure
of the entanglement available as a resource for quantum
information purposes \cite{regular}.

The symmetry of the states dramatically simplifies the calculation
of the relative entropy of entanglement. As follows immediately
from the convexity of the relative entropy and the invariance
under joint unitary operations, the closest PPT state can always
be taken to be a state of the same symmetry \cite{Rains,KGH}.
Hence, the closest PPT state is characterized by the same small
number of parameters. For simplicity of notation, we will denote
the subset of state space corresponding to specific numbers
$\alpha$, $\beta$ of photons as $(\alpha,\beta)$-photon space. In
the $(1,1)$-photon space let us denote the closest PPT state as
\begin{eqnarray}
\sigma^{(1,1)}= \zeta^{(1,1)}_{0} \Omega^{(1,1)}_{0} +
(1-\zeta^{(1,1)}_{0}) \Omega^{(1,1)}_{1}. 
\end{eqnarray}
Forming the partial
transpose of this state, and demanding that the 
resulting operator be non-negative, gives the condition  $
\zeta^{(1,1)}_{0} \leq 1/2$. In this simplest space, all symmetric
states lie on the straight line segment $\mu^{(1,1)}_{0} \in
[0,1]$ with the PPT region extending from the origin to the
midpoint (see Fig.\ 1).

In general, for higher photon numbers $\alpha$ and $\beta$, the
set of symmetric states are represented by a simplex in a
$(\text{min}(\alpha,\beta)+1)$-dimensional space, the coordinates
of which are denoted by $\mu^{(\alpha,\beta)}_j$. In turn, the PPT
criterion gives rise to a number of linear inequalities, such that
the set of invariant operators with a positive  partial transpose
corresponds again to a simplex. The intersection of the two
simplices corresponds to the invariant PPT states, and the
coordinates are denoted by  $\zeta^{(\alpha,\beta)}_j$
\cite{Hendriks}.

The situation with $\alpha=\beta =1,2,3$ is depicted explicitly in
Fig.\ 1. The simplex corresponding to symmetric states,
characterized by the condition that the $ \mu^{(\alpha,\beta)}_j$
form a probability distribution, is in these three cases a
straight line segment, an equilateral triangle, and a regular
tetrahedron respectively. The vertices of the simplex represent
the normalized projectors $\Omega^{(\alpha,\beta)}_{j}$. States in
the interior of the simplex are convex combinations of all the
allowed projectors. The PPT set with the same symmetry is clearly
marked.

\begin{figure}[t]\label{triangles}
\includegraphics[width=2.9 in]{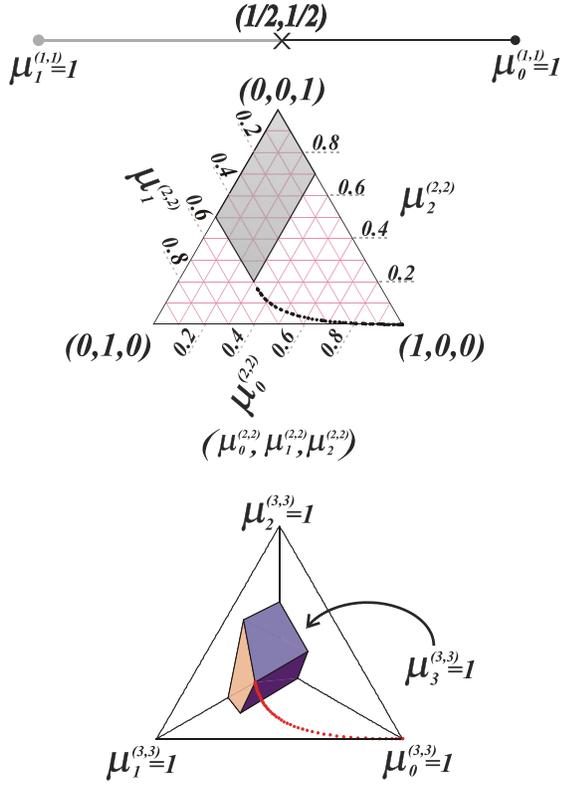}
\caption{\small{  The simplices of symmetric states for the cases of
$(\alpha,\beta)$, $\alpha=\beta=1,2,3$, respectively.  The
equilateral triangle has been marked with contour lines, on each
of which one of the parameters is constant.  The set of PPT states
is indicated by the grey line segment in the top graph, the shaded
area of the $(2,2)$ triangle and the filled polygon which obscures
the $\mu^{(3,3)}_3=1$ vertex in the $(3,3)$ tetrahedral space. In
all graphs only the projector of highest spin is within the PPT
set. For all three cases, the set of all possible down-conversion
states is a curve ending at the boundary of the PPT set, shown by
the solid black line for the (1,1) space, and by the dotted curves
for the (2,2) and (3,3) spaces. The position of the state on the
curve is determined by the parameter $\xi=(1-\eta)\tanh (\kappa t)$.}}
\end{figure}

\begin{figure}[t]\label{entanglement}
\includegraphics[width=0.79 \columnwidth]{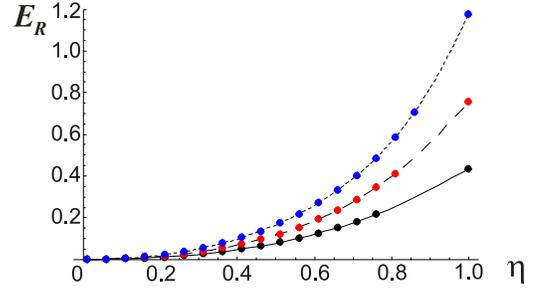}
\caption{\small{Lower bounds to the relative entropy of
entanglement for down-conversion states with initial average
photon numbers of $0.5$ (solid), $1$ (dashed), and $3$ (dotted line)
subject to loss, evaluating
the sum of Eq.\ (\ref{full}) up to a truncation of $\alpha,\beta\leq 5$.
This gives a
good approximation to the total entanglement for average photon
numbers before loss up to about $3$.  }}
\end{figure}

Fig.\ 1 also shows the curves traced by the down-conversion states
when they are subject to loss. As discussed above, the position of
the states on the curve is determined by the single parameter
$\xi$. For perfect transmission corresponding to $\eta=1$ the
quantum state in an $\alpha=\beta$ photon space has
$\mu^{(\alpha,\beta)}_{0}=1$ for all values of $t$, corresponding
to maximal entanglement. As losses are increased the state
migrates   towards the PPT boundary. 
It is an important immediate consequence of Eq.\ (\ref{muzero})
that for all losses $\eta>0$, the number $\mu^{(\alpha,\alpha)}_{0}$ is 
always greater than $1/2$ for any finite $t$ and for all $\alpha$. 
For any finite $t$,
$\mu^{(\alpha,\alpha)}_{0}\rightarrow 1/2$ as $\xi\rightarrow 1$ (which 
corresponds to 
a  limit of zero transmission time 
and infinite interaction
time). This holds true for $(\alpha,\alpha)=(1,1)$, but also for 
higher values of $\alpha$: the state
remains outside the PPT set for any non-vanishing $t$ and for
arbitrarily high losses. Therefore, the above results show that there is
always some entanglement in the down-conversion state, as quantified in 
terms of the relative entropy of entanglement. 
As a corollary, 
which one can already infer from the lowest dimensional subspace,
$(\alpha,\alpha)=(1,1)$,
there is actually distillable entanglement in the down-conversion
state, regardless of how lossy the transmission from the source to the
detector.

We now proceed to quantify the entanglement in the states more
explicitly. Since $E_R$ is convex and the set of symmetric PPT
states is convex, finding the closest state $\sigma$ amounts to
solving a convex optimisation problem. For different values of
$\alpha,\beta$ the quantities
$S(\rho^{(\alpha,\beta)}||\sigma^{(\alpha,\beta)})$ have been
evaluated, where $\sigma^{(\alpha,\beta)}$ denotes the
PPT state which is the unique global minimum in the convex
optimization problem, i.e., 
the PPT state closest to the down-conversion state.
For generic states, this optimization problem would still
be convex, yet, the dimensionality of state space grows as
$(\alpha+1)^2(\beta+1)^2-1$. The symmetry dramatically reduces the
dimensionality of the constraint set to searched
to $\text{min}(\alpha,\beta)$, and thus makes the quantification of
the entanglement a feasible task. For instance, for a state with
three photons on each side, one has to consider only three
objective variables instead of 255. The total relative entropy of
entanglement is given by the expression
\begin{eqnarray}\label{full}
E_{R}(\rho) = \sum_{\alpha,\beta=0}^{\infty} P{(\alpha,\beta)}
E_{R} (\rho^{(\alpha,\beta)}).
\end{eqnarray}
The average photon number
before loss $N$ is related to the interaction time $t$ as $N=2
\sinh^2 (\kappa t)$. The average photon number after loss is $n=\eta
N$. Fig.\ 2 shows
the relative entropy of entanglement
calculated as described above for $N=0.5$, $N=1$ and $N=3$. One
sees that significant entanglement remains even for substantial
losses.

\section{Conclusions}

We have shown how symmetry considerations make possible the
quantification of entanglement for states produced by parametric
down-conversion and subject to losses. The resilience of the
entanglement of these multi-photon states under photon loss makes
them an excellent system for the experimental demonstration of
entanglement of large photon numbers \cite{eisenberg} and good
candidates for quantum communication schemes \cite{durkin}.

\begin{acknowledgements}

G.A.D.\ was supported by EPSRC, GR/M88976, J.E.\ by the European
Union (EQUIP, IST-1999-11053 and QUPRODIS, IST-2001-38877), the
A.-v.-Humboldt Foundation and the DFG (Schwerpunkt QIV),
and C.S.\ by a Marie Curie
Fellowship of the EU (HPMF-CT-2001-01205).

\end{acknowledgements}


\begin{references}
\bibitem{Int}
    The measures distillable entanglement and entanglement cost
    specify certain optimal conversion rates: these are the rates that
    can be achieved in an asymptotic extraction of and preparation
    procedure starting from maximally entangled qubit pairs. 
    These procedures are thought to be implemented by employing
    local quantum operations and classical communication (LOCC)
    only \cite{Bennett,Rains}.
\bibitem{regular}
	With the methods of Refs.\ \cite{Regul}, the asymptotic
	versions $E_\infty (\rho)= \lim_{n\rightarrow \infty} 
	E_R(\rho^{\otimes n})/n$ would in principle 
	also be accessible for this
	class of states.
\end{references}
\end{document}